\documentclass[aps,prl,reprint,groupedaddress]{revtex4-1}
\usepackage[dvipdfmx]{graphicx}

\begin{document}


\title{Exciton splitting in semiconducting carbon nanotubes in ultrahigh magnetic fields above 300 T}


\author{Daisuke Nakamura$^1$, Tatsuya Sasaki$^1$, Weihang Zhou$^1$, Huaping Liu$^2$, Hiromichi Kataura$^3$, and Shojiro Takeyama$^1$}
\affiliation{$^1$Institute for Solid State Physics, University of Tokyo, 5-1-5 Kashiwanoha, Kashiwa, Chiba 277-8581, Japan\\
$^2$Beijing National Laboratory for Condensed Matter Physics, Institute of Physics, Chinese Academy of Sciences, Beijing 100190, China\\
$^3$Nanosystem Research Institute, National Institute of Advanced Industrial Science and Technology, Tsukuba, Ibaraki 305-8562, Japan}


\date{\today}

\begin{abstract}
In high magnetic fields, the exciton absorption spectrum of a semiconducting single-walled carbon nanotube splits as a result of Aharonov-Bohm magnetic flux.
A magnetic field of 370 T, generated by the electro-magnetic flux compression destructive pulsed magnet-coil technique, was applied to single-chirality semiconducting carbon nanotubes. 
Using streak spectroscopy, we demonstrated the separation of the independent band-edge exciton states at the K and K' points of the Brillouin zone after the mixing of the dark and bright states above 150 T. 
These results enable a quantitative discussion of the whole picture of the Aharonov-Bohm effect in single-walled carbon nanotubes.
\end{abstract}


\maketitle


\section{Introduction}

The penetration of magnetic flux quanta into a material breaks the time-reversal symmetry of the electronic states and lifts degeneracies of charge, spin, and orbital degrees of freedom. 
However, eliciting the effect in nanometer-size substances requires very strong magnetic fields. 
Rolled graphene with a certain chirality ($n$, $m$) forms a single-walled carbon nanotube (SWNT). 
The typical diameter of an SWNT is 1 nm; the penetration of one flux quantum ($\phi =\phi_0$) into such a nanotube, producing one period of the band-edge oscillation as a result of the Aharonov-Bohm (A-B) effect, requires an external magnetic field of 5200 T. 
As an ideal material for investigating the A-B effect, the magneto-optical properties of SWNTs have been extensively studied theoretically.\cite{Ajiki1994, Ando2004, Spataru2004, Ando2006, Capaz2006, Ajiki2012}

The effects of exciton and exchange interactions are pronounced, even at room temperature, and govern the band-edge optical properties in SWNTs.
Exciton spectra have been used for the characterization of SWNTs, especially for the assignment of chirality.\cite{Bachilo2002}
There are 16 quantum states identified as excitons at the band-edges, but only one, known as the zero-momentum singlet exciton (hereafter called the bright exciton) is optically active. 
Owing to the A-B effect, a magnetic flux, $\phi$, penetrating the cross section of an SWNT induces valley splitting between the K and K' points in the Brillouin zone, along with a mixing of the exciton wave functions. 
Therefore, the bright exciton and its dark counterpart are known to convert into two bright K and K' point excitons upon application of a magnetic field parallel to the tube axis, hereafter denoted as $B_{\text{eff}}$.\cite{Ando2004}
According to theory within the $k\cdot p$ scheme presented by Ando,\cite{Ando2006} the absorption spectra of these K and K' point excitons evolve differently in high magnetic fields, depending on the relative energy-level ordering of the band-edge bright and dark excitons (Fig. 1). 
However, because of the nanometer scale of the tube diameter ($d_t$), an ultrahigh magnetic field with a strength far beyond that available in conventional laboratories is required for capturing the whole picture of the A-B effect in SWNTs. 

\begin{figure}
\includegraphics[width=8.5cm]{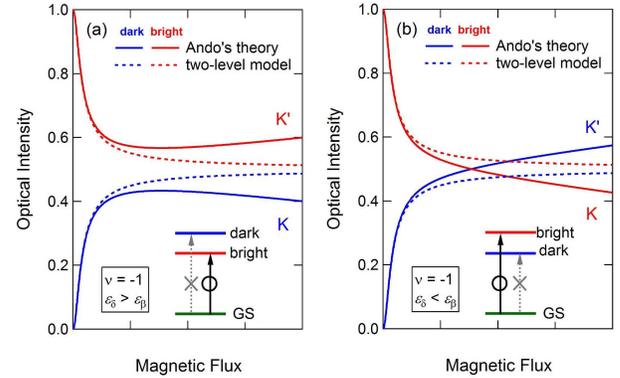}%
\caption{\label{fig1} Schematic evolution of the absorption intensity as a function of the A-B flux for SWNTs with the family pattern $\nu$ = -1. 
In case that the bright exciton energy is (a) smaller and (b) larger than that of dark exciton.
Solid curves indicate Ando's theory,\cite{Ando2006} and dashed curves are the predictions of the two-level model.}
\end{figure}

\begin{figure*}
\includegraphics[width=17cm]{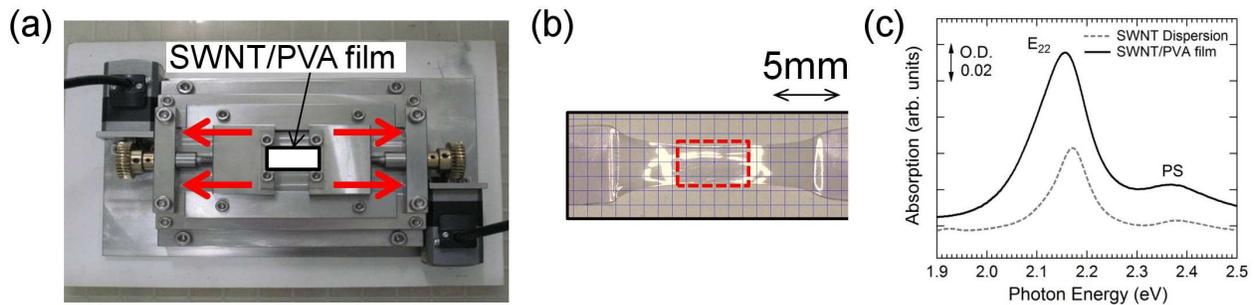}%
\caption{\label{fig3} (a) Laboratory-built stretching apparatus for the SWNT-embedded PVA film. 
(b) Photograph of the stretched SWNT/PVA film. 
For magneto-optical measurement, only the uniformly stretched region was used (enclosed by the dashed line).
(c) Absorption spectrum of (solid curve) the stretched film of SWNT/PVA composite and (dashed curve) SWNTs dispersed in a liquid of 0.05 wt \% DOC, which was measured before the synthesis of the SWNT/PVA composite film. 
A phonon sideband (PS) peak was observed on the higher energy side of the main $E_{22}$ exciton absorption peak. }
\end{figure*}

Owing to the quasi-one-dimensional tubular structure of nanotubes, the electron wave vector in the circumferential direction is quantized and produces a set of sub-bands. 
The first sub-band ($E_{11}$) transition takes place in the near-infrared optical region, and the second sub-band ($E_{22}$) transition remains in the visible light region in most of the available SWNTs. 
So far, in attempts to study A-B effects in SWNTs, both absorption and photoluminescence (PL) measurements for the $E_{11}$ transition have been performed by many groups.\cite{Arnold2004, Zaric2004, Dukovic2005, Wang2005, Zaric2006, Mortimer2007PRB, Mortimer2007PRL, Shaver2007, Srivastava2008, Matsunaga2008, Yokoi2010, Takeyama2011PRB}
In contrast, there have been very few systematic investigations of the $E_{22}$ transition, owing to a broader absorption peak and the lack of PL. 
However, probing the $E_{22}$ transition is necessary for the investigation of exciton spectra above 200 T, because of technical difficulties in carrying out the magneto-absorption measurement called "streak spectroscopy" of infrared region (corresponding to $E_{11}$ transition) in such an ultra-high magnetic field environment.

Our group previously made an attempt to capture the whole picture of the A-B effect in the $E_{22}$ exciton up to 300 T by using a D$_2$O solution of SWNTs grown by the high-pressure carbon monoxide (HiPco) method.\cite{Takeyama2011JPCS}
However, the mixed chirality of the SWNTs in a sample obscured the peak splitting because peaks with different kinds of chirality overlapped at very high magnetic fields. 
Furthermore, $B_{\text{eff}}$ was substantially reduced owing to the random orientation of individual SWNTs dispersed in the liquid. 
Therefore, two conditions must be fulfilled to avoid the above problems: (i) the use of a sample with a single chirality without any background absorption, and (ii) the preparation of highly aligned SWNTs, as only $B_{\text{eff}}$ contributes to the A-B flux. 
However, when we used highly aligned single chirality SWNTs, the spectra of the $E_{22}$ exciton were not sufficiently split, even for a magnetic field as strong as 186 T;\cite{Zhou2013PRB, Zhou2014SRep} therefore, the whole picture of the A-B effect in SWNTs predicted by Ando\cite{Ando2004} was not quantitatively discussed. 
In this paper, we present results from streak spectroscopy measurements of the $E_{22}$ transitions performed for highly aligned single chirality SWNTs in magnetic fields up to 370 T. 

\section{Experiment}

We prepared highly isolated (6,5) SWNTs from a mixture of HiPco-grown carbon nanotubes by single-surfactant multicolumn gel chromatography, which was recently developed by some of the authors.\cite{Liu2011} 
The SWNTs were dispersed in a 0.05 wt\% sodium deoxycholate (DOC) solution, mixed with a water solution with 15 wt\% polyvinyl alcohol (PVA), and naturally dried for 2-3 days to make a film. 
The SWNTs in the PVA film were aligned by stretching the film with a laboratory-built stepping-motor-driven stretching apparatus (Fig. 2(a)). 
After heating to 80 - 100 $^\circ$C with an infrared heat lamp, the film was stretched at a constant speed of 0.33 mm/s, and a final stretching ratio of 5 was obtained. 
For the magneto-optical measurement, only the uniformly stretched region was used (the area enclosed by the dashed line in Fig. 2(b)).

The ensemble-averaged angle between the axes of the SWNTs and the stretching direction (corresponding to the direction of the external magnetic field), $\theta_{\text{ave}}$, was evaluated from the optical anisotropy, $A$, along the stretching direction of the SWNT-embedded film, via the nematic order parameter, $S$ [$S = (3 \cos 2\theta_{\text{ave}} -1)/2 = (1-A)/(1+2A)$ ].\cite{Takeyama2011PRB} 
Optical anisotropy measurements using a spectrophotometer gave $S$ = 0.66, yielding $\theta_{\text{ave}}$ = 29$^\circ$.\cite{Zhou2013PRB} 
Therefore, the effective magnetic field $B_{\text{eff}} = B \cos \theta_{\text{ave}}$ along the averaged direction of the stretched SWNTs was determined to be 370 $\times \cos$ 29$^\circ$ = 320 T at maximum in this study, corresponding to $\phi$ = 0.035$\phi_0$ for (6,5) SWNTs possessing a tube diameter of $d_t$ = 0.757 nm. 

We note that the $E_{22}$ peak position of the stretched film is red shifted by 15 meV compared with that of an SWNT dispersion before the synthesis of the SWNT/PVA composite film (Fig. 2(c), overall spectrum is shown in Ref. 22). 
This is known as the environment dielectric screening effect on the SWNT exciton, which has been widely discussed.\cite{Derycke2002APL, OConnell2002Science, Zhang2003NLett, Li2004PRL, Ando2010, Fagan2013}
According to the calculation by Ando,\cite{Ando2010} the exciton energy decreases with increasing the dielectric constant of surrounding material.
For SWNT dispersion in Fig. 2(c), there is a hydrophobic surfactant layer of DOC around SWNT, which disturbs the environmental effect from the water molecule \cite{Fagan2013}.
On the other hand, for SWNT-embedded PVA film, we consider that some water molecules exist near SWNT due to the high moisture sensitivity of PVA, and high dielectric constant of water (80.4 at 20 C$^\circ$) induces the significant dielectric screening effect on the SWNT exciton spectra.
Similar redshift of exciton peak position in SWNT-embedded PVA film is reported by Zhang.\cite{Zhang2003NLett}

\begin{figure}
\includegraphics[width=7cm]{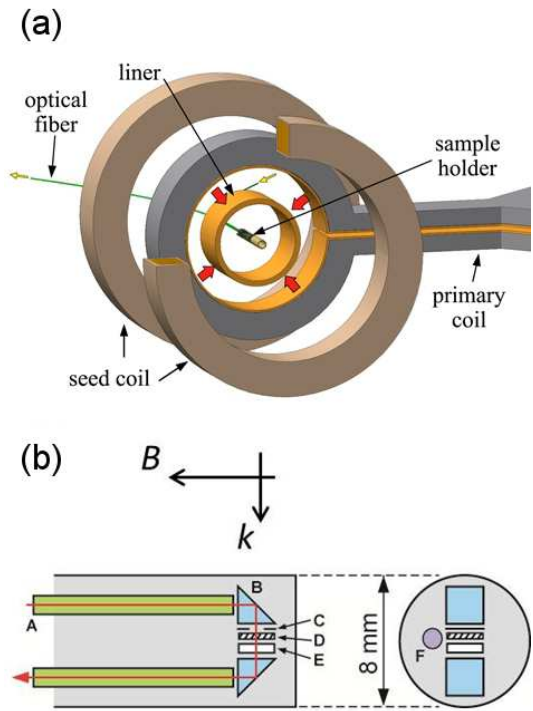}%
\caption{\label{fig2} (a) Illustration of the EMFC magnet with a sample holder, which is placed at the center of the imploding liner. 
(b) Schematic view of the detailed structure of the sample holder. 
A: optical fiber; B: prism; C: mask; D: polarizer; E: sample; F: magnetic field pickup coil. }
\end{figure}

Ultrahigh magnetic fields above 300 T can only be achieved by a destructive magnetic flux-compression method employing extremely short pulses with an order of a microsecond. 
In the present study, the electro-magnetic flux compression (EMFC) technique was employed. 
A magnetic field intensity of around 700 T is the maximum value that can be used for solid-state physics measurements with sufficiently high accuracy and reliability.\cite{Takeyama2011JPD} 
A schematic illustration of the EMFC magnet is shown in Fig. 3(a). 
High-density magnetic flux is achieved by an imploding metal cylinder, called the "liner," rapidly compressing an initial magnetic flux of the order of 4 T that is generated by a pair of seed field coils. 
A huge electrical current of several mega-amperes is injected from the condenser bank into the primary coil, which induces the electromagnetic force for imploding the liner. 
In this experiment, the condenser banks provided 4.5 MJ of energy. 
The inner radius of the liner shrinks to a few millimeters at the peak of the magnetic field, and decreases as the maximum magnetic field intensity increases (shown in Fig. 8 of Ref. 30). 
Therefore, for solid-state physics experiments using the EMFC technique, the outer diameter of the sample holder limits the maximum available magnetic field owing to collision with the imploding liner before reaching the peak field. 

In this study, a small sample holder with a diameter of 8 mm was designed while maintaining sufficient optical transmission intensity (Fig. 3(b)). 
The sample holder, made from a bakelite (phenol resin) tube, accommodates a polarizer and the sample film sandwiched by a pair of prisms to configure the optical $k$-vector perpendicular to the $B$-axis (the Voigt geometry). 
The axis of the polarizer is parallel to the stretching direction of an SWNT film. 
As there is significant electromagnetic noise and flashing light from the imploding liner, a super-insulation foil is wound outside the sample holder for protection.\cite{Nakamura2013RSI}
A Xe flash arc lamp was used as the light source. 
By using a streak camera (Hamamatsu Photonics, C4187-25S), the magneto-absorption spectra of SWNTs were continuously measured while generating the pulsed magnetic field. 
All spectra were acquired at room temperature.

\section{Results and Discussion}

\begin{figure*}
\includegraphics[width=14cm]{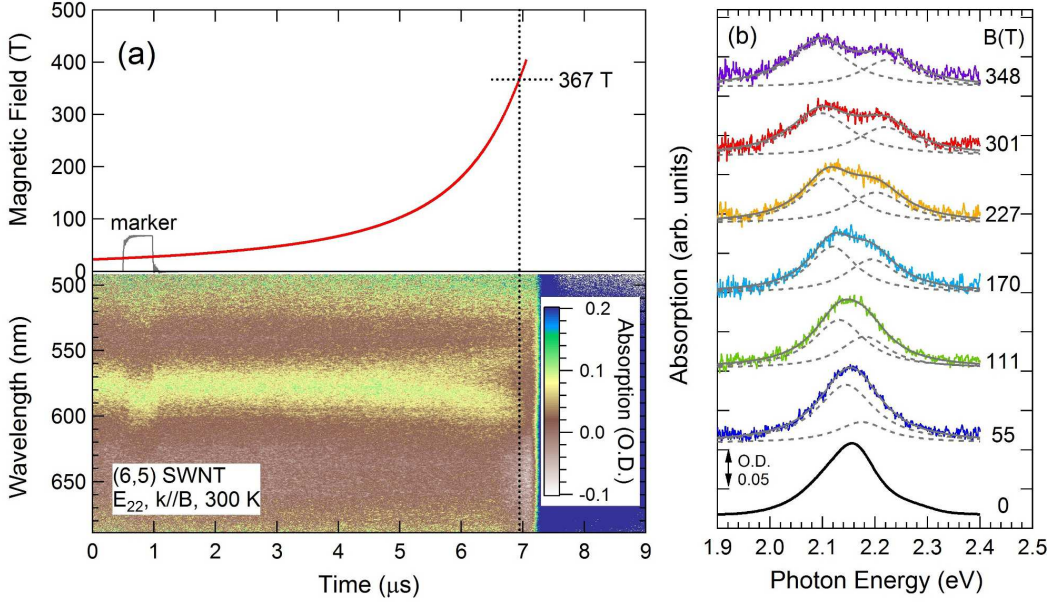}%
\caption{\label{fig4} 
(a) (upper panel) Curve showing the magnetic field evolution in time generated by the EMFC technique. 
(bottom panel) Streak image synchronized with the magnetic field. 
(b) Magneto-absorption spectra for various magnetic fields. 
Dashed curves are the results of Lorentzian curve deconvolution applied to each absorption peak.
The bottom solid curve shows the absorption spectrum at $B$ = 0, taken by the spectrophotometer.
}
\end{figure*}

Figure 4(a) shows the magnetic field curve (top panel) and a two-dimensional streak image of the optical absorption of the semiconducting (6,5) SWNTs (bottom panel) detected by a streak camera synchronized with the pulsed magnetic field. 
The magneto-absorption spectrum and the magnetic field curve disappear at almost the same time ($\sim$ 7$\mu$s) owing to the destruction of the sample holder in the final stage of magnetic field generation. 

In Fig. 4(b), typical magneto-absorption spectra of $E_{22}$ excitons for various magnetic field strengths were extracted from Fig. 4(a) and compared with that measured by the spectrophotometer at $B$ = 0. 
The phonon sideband peak (denoted as PS in Fig. 2(c)) located on the higher energy side of the main $E_{22}$ exciton absorption peak was subtracted in Fig. 4(b), as we have confirmed that the phonon sideband peak does not change with magnetic field up to 52 T.\cite{Zhou2013APL} 
A single exciton peak located at 2.155 $\pm$ 0.005 eV splits into two with almost identical intensities at the highest magnetic field. 
The evolution of the split peaks clearly demonstrates that a new peak emerges on the higher energy side of the bright exciton peak, which is consistent with the magneto-absorption experiment recently carried out up to 186 T using the single-turn coil (STC) technique.\cite{Zhou2013PRB, Zhou2014SRep}

The evolution of the peak energies and intensities of the splitting absorption spectra were tracked by deconvolution using Lorentzian functions (dashed curves in Fig. 4(b), see also Appendix Sec. I), and are summarized in Fig. 5. 
The energy ($\varepsilon_{\beta, \delta}$) and the integrated intensity ($I_{\beta, \delta}$) of each peak for the bright ($\beta$) and dark ($\delta$) excitons are plotted as a function of $B_{\text{eff}}$ (squares and circles: the EMFC experiments; triangles: the previous STC experiment\cite{Zhou2013PRB}). 
The overall features of the intensity and energy evolution shown in Fig. 5 are quite consistent with the calculations provided by Ando\cite{Ando2006} for the case that the bright exciton is located on a lower energy at the band-edge shown in Fig. 1(a). 
One difference is the rather small exciton splitting energy, $\Delta_{AB} = \mu B_{\text{eff}}$ ($\mu$: exciton splitting coefficient), which is also observed for $E_{11}$ excitons.\cite{Zhou2013PRB}
For comparison, the theoretical splitting for a given energy band gap in the absence of interaction, which is given by $\mu_{th} = 3(\pi ed_t^2 /2h) E_g$ ($E_g = 4\pi \gamma /3L$: energy band gap in the absence of interaction),\cite{Ajiki1994} is shown in Fig. 5(a) as dashed curves. 
For (6,5) SWNTs, $\mu_{th}$ = 0.74 meV/T.

\begin{figure}
\includegraphics[width=8.5cm]{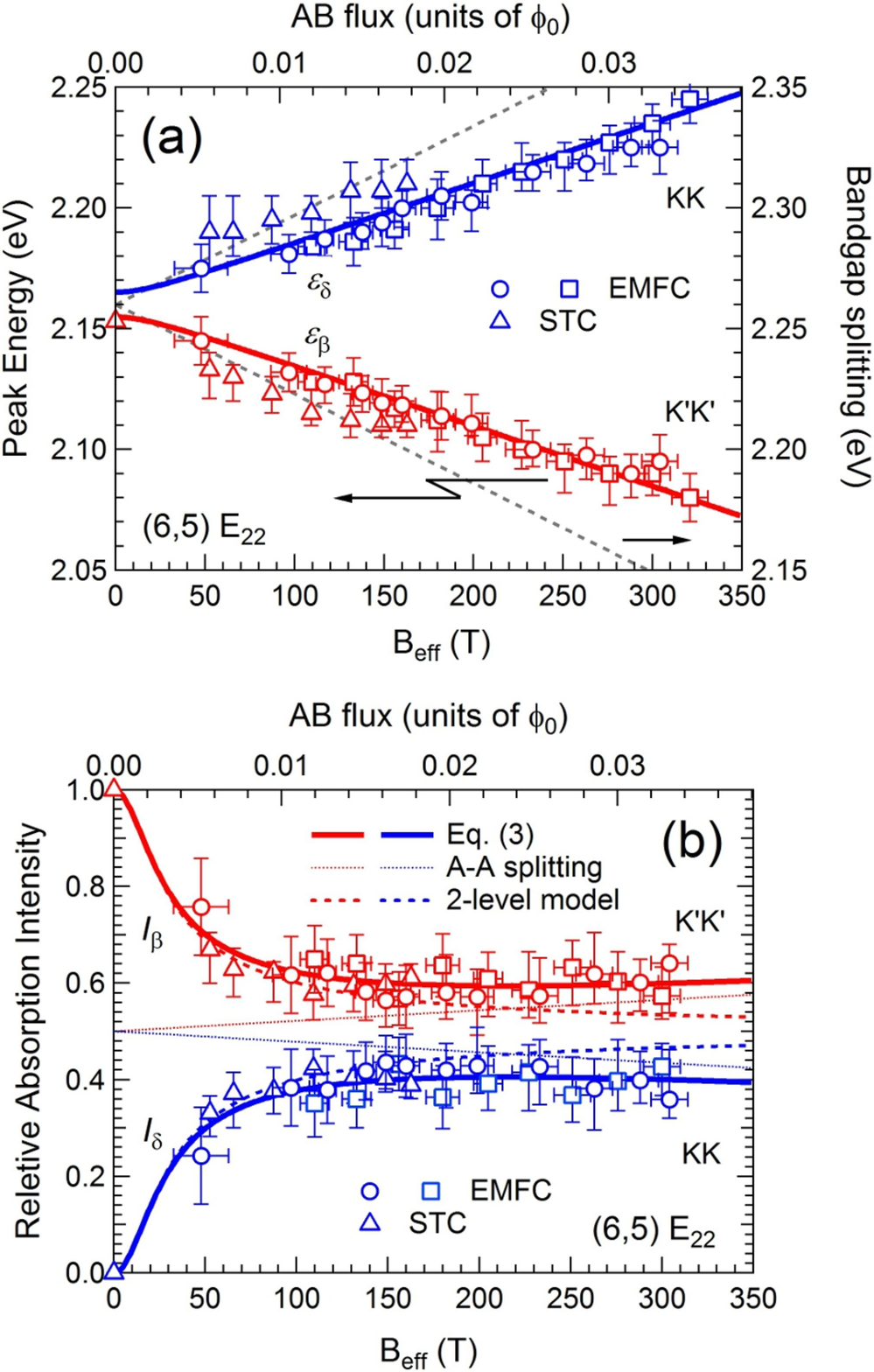}%
\caption{\label{fig5} (a) Energy and (b) integrated intensity as a function of the effective magnetic field. 
Bold curves are fitting results. 
Data denoted by STC are taken from Ref. 21. 
In (a), the splitting for a given energy band gap is indicated by dashed curves. 
In (b), thin dotted curves depict the A-A splitting term in Eq. (3), and bold dashed curves are the two-level model predictions.}
\end{figure}

According to the Ando's theory \cite{Ando2006} shown in Fig. 1, for $\varepsilon_\beta < \varepsilon_\delta$, the absorption intensities of dark and bright excitons decrease and increase, respectively, in proportion to the magnetic flux in the high field limit. 
This is because the mixing between the K and K' points diminishes with increasing magnetic flux, and thus excitons at the K and K' points become independent of each other. 
However, owing to the lack of sufficient A-B flux in SWNTs, previous magneto-optical experiments utilized the following approximate formula (two-level model\cite{Shaver2007}, dashed curves in Fig. 1) for analyzing their results.
\begin{equation}
 I_{\beta,\delta} = \frac{1}{2} \pm \frac{1}{2} \frac{\Delta_{bd}}{\sqrt{\Delta_{bd}^2 + \Delta_{\text{AB}}^2 (B_{\text{eff}})}}
\end{equation}
\begin{equation}
 \varepsilon_{\beta,\delta} = \frac{\varepsilon_\beta (B_{\text{eff}} = 0) + \varepsilon_\delta (B_{\text{eff}} = 0) }{2} \pm \frac{\sqrt{\Delta_{bd}^2 + \Delta_{\text{AB}}^2 (B_{\text{eff}})}}{2}
\end{equation}
where $\Delta_{bd}$ represents the zero-field splitting between the bright and dark excitons. 
The two-level model is insufficient to describe the split spectra of unmixed K and K' point excitons, since $I_\beta$ converges to $I_\delta$ in the high field limit, which is different from Ando's theory\cite{Ando2006}. 
In Fig. 5(b), $I_\beta$ and $I_\delta$ tend to repel each other as the magnetic field increases above 150 T, indicating a limit of the two-level model. 
As the peak intensity of the dynamical conductivity given by the perturbation method is inversely proportional to the exciton energy,\cite{Ando2004} a term linear in $B$ is added to Eq. (1). 
For $\varepsilon_\beta < \varepsilon_\delta$, the resultant evolution of intensity with $B_{\text{eff}}$ is described by 
\begin{equation}
 I_{\beta,\delta} = \frac{1}{2} \pm \frac{1}{2} \frac{\Delta_{bd}}{\sqrt{\Delta_{bd}^2 + \Delta_{\text{AB}}^2 (B_{\text{eff}})}} \pm \alpha \frac{\pi d_t^2}{4\phi_0}B_{\text{eff}}
\end{equation}
where $\alpha$ is a dimensionless coefficient. 
In Fig. 5(b), the bold curves are the results of fitting using Eq. (3). 
The thin dotted curves in Fig. 5(b) show the term linear in $B$ in Eq. (3), hereafter called the Ajiki-Ando (A-A) splitting term. 
For the best fit, the parameters are $\mu_{ex}$ = 0.50 $\pm$ 0.05 meV/T, $\Delta_{bd}$ = 10.3 $\pm$ 1.5 meV, and $\alpha$ =1.0 $\pm$ 0.2. 
We confirmed that the evolution of the absorption intensity for the $E_{11}$ exciton in the same sample\cite{Zhou2013PRB} was also well-fitted by using the same value of $\alpha$ = 1.0 (see Appendix Sec. II). 

\begin{figure}
\includegraphics[width=7cm]{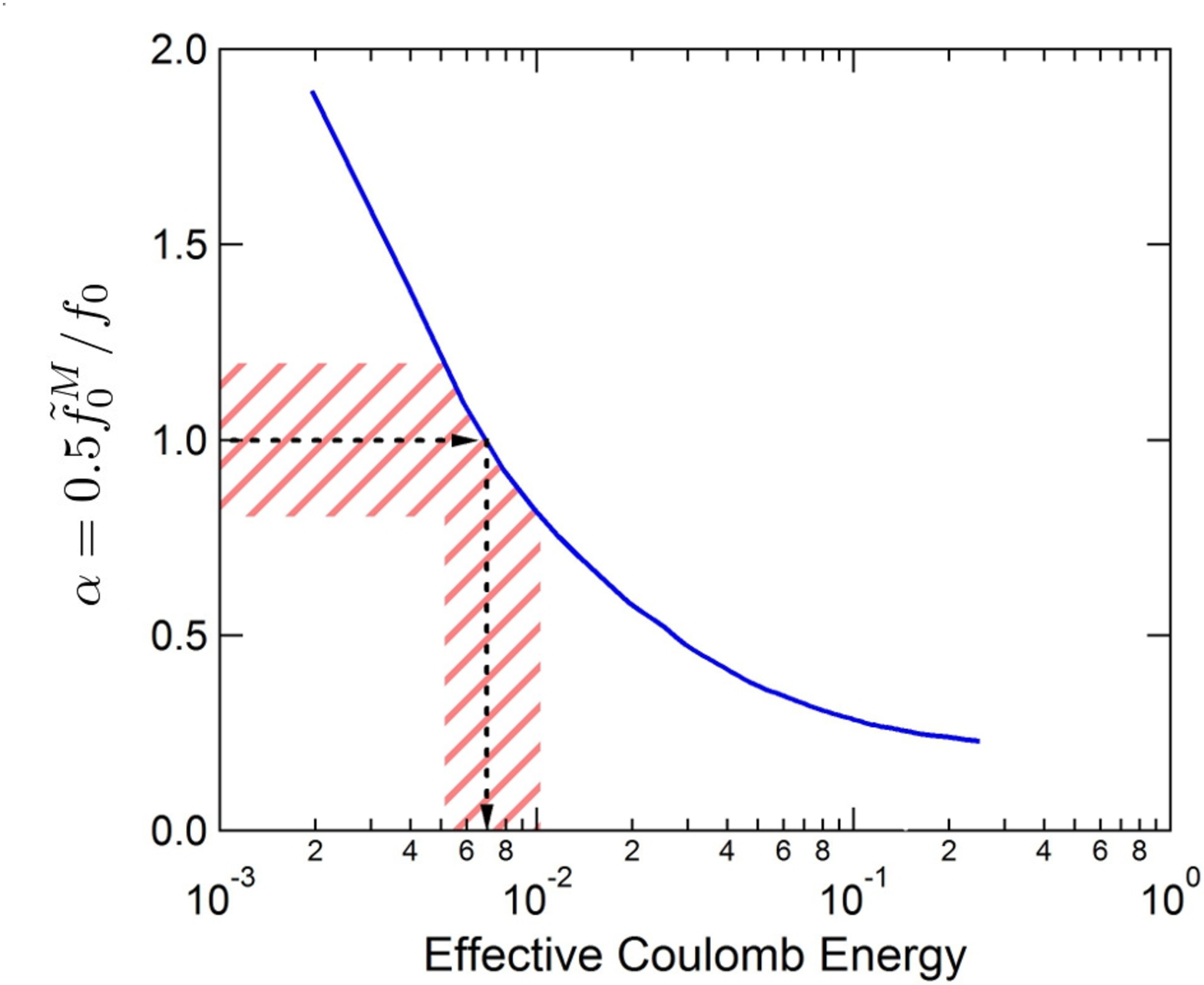}%
\caption{\label{fig7} Coefficient $\alpha = 0.5 \tilde f_0^M / f_0$  as a function of the effective Coulomb energy for $d_t$ about 0.7 nm using Ando's calculation.\cite{Ando2004}
Dotted arrows indicate the value of the effective Coulomb energy determined by the present study, and the hatched region shows the estimated error-bar.}
\end{figure}

According to the formularization by Ando\cite{Ando2004}, the magnetic field dependence of the oscillator strength of unmixed K and K' point excitons is described as a function of the A-B flux $\phi$:
\begin{eqnarray}
  \left\{
    \begin{array}{l}
    f_0^K (\phi) = f_0 - \tilde f_0^M \phi  \\
    f_0^{K'} (\phi) = f_0 + \tilde f_0^M \phi
    \end{array}
  \right.
\end{eqnarray}
Here, $f_0$ is the oscillator strength of the K and K' point excitons at zero magnetic field and $ \tilde f_0^M$ is the linear coefficient of the A-B flux. 
Therefore, $\alpha$ is equivalent to  $0.5 \tilde f_0^M / f_0$ in the formularization given by Ando. 
In Fig. 6,  $0.5 \tilde f_0^M / f_0$ is plotted for $d_t$ of about 0.7 nm as a function of effective Coulomb energy, $(e^2/\kappa L)(2\pi \gamma /L)^{-1}$ ($\kappa$: static dielectric constant for states except those lying near the Fermi level and for surrounding materials; $\gamma$: band parameter; $L$: circumference).\cite{Ando2004}
Here, $\mu_{ex}$ is employed instead of $\mu_{th}$. 
Following the dotted arrow in Fig. 6, the present value of $\alpha$ = 1.0 corresponds to $(e^2/ \kappa L)(2 \pi \gamma /L)^{-1}$ of around 0.007, and leads to a value of $\kappa$ around 50.
The static dielectric constant in semiconducting SWNTs is reported typically less than 5;\cite{Krupke2003} however, a value of $\kappa$ can easily become high as 50 by the environmental material.

The redshift by 15 meV (0.7 \%) of the $E_{22}$ exciton peak relative to that of the original SWNT dispersion observed in Fig. 2(c) is another evidence of the existence of a significant environmental effect.
The value of $\kappa \sim$ 50 obtained here leads to the redshift in the $E_{22}$ exciton peak estimated as to be $\sim 0.6 \%$ for $d_0 = 2d_t$ and $\sim 1.9 \%$ for $d_0 = 1. 5d_t$, deduced from Fig. 4 in Ref. 28.
Here, $d_0$ is the inner diameter of a dielectric medium surrounding the SWNT.
For (6,5) SWNT, $d_0/d_t$ becomes 1.88, if $(d_0-d_t)/2$ is set as the interlayer distance of graphite (3.34 $\AA$).\cite{Ando2010}
Therefore, 0.7 \% redshift of the $E_{22}$ exciton peak shown in Fig. 2(c) is a reasonable value when SWNT is affected by highly dielectric material such as residual water molecules in hydrophilic PVA film, and is consistent with the magnitude of A-A splitting term ($\alpha$ = 1.0).

In conclusion, in a semiconducting (6,5) SWNT, the A-B magnetic flux causes quantum state mixing of band-edge dark and bright excitons, and the optical activity (oscillator strength) is shared by both of them to produce almost equal absorption intensity. 
As the magnetic field rises, these states split into independent excitons at the K and K' points, respectively, and show a simple energy splitting that is linear in magnetic field strength. 
For $E_{22}$ exciton, the oscillator strength of the K' point exciton increases, whereas that of the K point exciton decreases linearly with increasing magnetic field strength. 
Owing to highly aligned stretched film of single chirality SWNTs, discernible splitting of the absorption spectra was observed above 300 T, which enabled us to conduct quantitative analysis of the A-B splitting of $E_{22}$ K-K' excitons.
 
\begin{acknowledgments}
H. Kataura acknowledges support by JSPS KAKENHI Grant Number 25220602. 
H. Liu acknowledges supports by the "100 talents project" of CAS and the recruitment program of global youth experts.
We thank Prof. H. Suzuura at Hokkaido University for fruitful discussions. 
We are grateful to Prof. S. Maruyama and Mr. S. Harish of the UTokyo for their kind instruction in sample preparation. We also thank Mr. H. Sawabe for his technical support for conducting the EMFC experiment.
\end{acknowledgments}

\bibliography{Ref.bib}

\appendix
\section{I. Deconvolution of the absorption peak of SWNT ensembles}
Since each SWNT follows the angular distribution function $f(\theta)$ along the magnetic field and the polarized light axes, the measured optical spectrum is a superposition of that from each SWNT. 
The function $f(\theta)$ is determined from $\theta_{\text{ave}} \equiv \int_0^{\pi/2} \theta f(\theta)d\theta/ \int_0^{\pi/2} f(\theta)d\theta$. 
Here, we postulated $f(\theta) = \exp [-(\theta/c)^2]$, and obtained $c$ = 0.94 for $\theta_{\text{ave}}$ = 29$^\circ$. 
Since the absorption intensity of a single SWNT ($I_{\text{single}}$) is proportional to $\cos ^2\theta$, the absorption spectrum is represented using the Lorentzian function as follows.
\begin{equation}
I_{\text{single}}(E,\theta) = \cos^2 \theta \times \frac{1}{\pi} \frac{a}{(E-E_0)^2+ a^2}
\end{equation}
$I_{\text{single}}(E,\theta_{\text{ave}})$ can be used in place of $I_{\text{ave}}$ (the superposition of $I_{\text{single}}(E,\theta)$), indicating that
\begin{equation}
I_{\text{ave}} = \frac{\int_0^{\pi/2} I_{\text{single}}(E,\theta) f(\theta)d\theta}{\int_0^{\pi/2} f(\theta)d\theta} \approx  I_{\text{single}}(E,\theta_{\text{ave}})
\end{equation}
which is shown in Fig. 7. 
The dashed curve is $I_{\text{ave}}$ and the solid curve is $I_{\text{single}}(E,\theta_{\text{ave}})$.
In the calculation, $a$ = 15 meV is used. 
Although the peak intensity of $I_{\text{ave}}$ decreases to 93 \% of that of $I_{\text{single}}(E,\theta_{\text{ave}})$, the full widths at half-maximum of the two curves are almost the same. 
Therefore, in Fig. 4(b), the sum of the two Lorentzian functions is used for fitting the magneto-optical spectrum and investigating the energy splitting and the relative intensity as a function of $B_{\text{eff}}$.

\begin{figure}
\includegraphics[width=8cm]{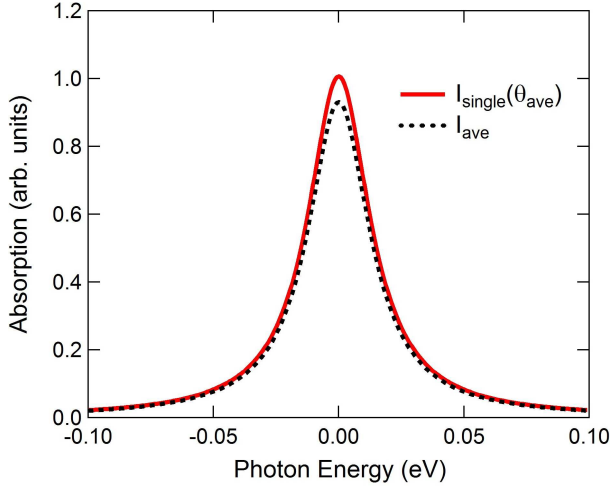}%
\caption{\label{figS1} Dashed curve is a superposition of the response from each SWNT ($I_{\text{ave}}$) follows the angular distribution function $f(\theta)$, and solid curve is $I_{\text{single}}(E,\theta_{\text{ave}})$.
}
\end{figure}

\section{II. Exciton splitting of the $E_{11}$ transition}
The evolution of the absorption intensity is predicted to be quite different for different relative orderings of the bright and dark excitons.\cite{Ando2006}
For $\varepsilon_\beta < \varepsilon_\delta$, a dark exciton changes to a K point exciton in the high field limit. 
On the other hand, for $\varepsilon_\beta > \varepsilon_\delta$, a bright exciton changes to a K point exciton. 
This is shown schematically in Fig. 1. 
When $\varepsilon_\beta > \varepsilon_\delta$, the sign of the A-A splitting term is reversed:
\begin{equation}
 I_{\beta,\delta} = \frac{1}{2} \pm \frac{1}{2} \frac{\Delta_{bd}}{\sqrt{\Delta_{bd}^2 + \Delta_{\text{AB}}^2 (B_{\text{eff}})}} \mp \alpha \frac{\pi d_t^2}{4\phi_0}B_{\text{eff}}
\end{equation}
To find the value of the coefficient $\alpha$, we can use experimental results for the $E_{11}$ transition of the same sample in this study, where $\varepsilon_\beta > \varepsilon_\delta$ is confirmed.\cite{Zhou2013PRB}
In Fig. 8, the optical absorption intensity data taken from Ref. 21 are fitted by Eq. (7). 
The fitting parameters are $\mu_{ex}$ = 0.57 meV/T, $\Delta_{bd}$ = 8.5 meV, and $\alpha$ = 1.0. 
The fitting functions (solid curves) almost perfectly describe the overall magnetic field dependence up to $B_{\text{eff}}$ = 110 T. 
We predict that the optical absorption intensities of bright and dark excitons become equal for the $E_{11}$ transition of (6,5) SWNTs at $B_{\text{eff}}$ = 185 T ($\phi / \phi_0$ = 0.02).
Therefore, the A-A splitting term is necessary for explaining the split spectra of unmixed K and K' point excitons in ultrahigh magnetic fields.

\begin{figure}
\includegraphics[width=8cm]{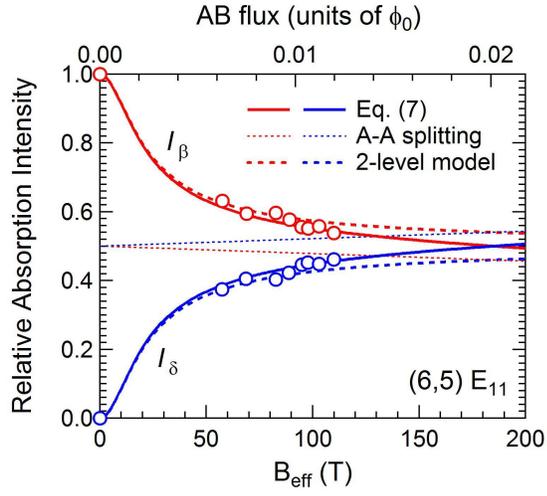}%
\caption{\label{fig8} Exciton splitting of the $E_{11}$ transition. 
Solid curves show the fitting result using Eq. (7), and the A-A splitting term in Eq. (7) is indicated by thin dotted curves. 
Thick dashed curves are the predictions of the two-level model for comparison. 
}
\end{figure}

\end{document}